\documentclass[fleqn,usenatbib]{mnras}

\usepackage{newtxtext,newtxmath}

\usepackage[T1]{fontenc}
\usepackage{ae,aecompl}

\usepackage{graphicx}	
\usepackage{amsmath}	
\usepackage{amssymb}	
\usepackage{ulem}

\newcommand{\dd}{\mathrm d}
\renewcommand{\d}{\partial}

\renewcommand{\(}{\left(}
\renewcommand{\)}{\right)}

\title[Particle acceleration at colliding shocks]{Particle acceleration at colliding shock waves}

\author[T. Vieu et al.]{
T. Vieu,$^{1}$\thanks{E-mail: vieu@apc.in2p3.fr}
S. Gabici,$^{1}$
V. Tatischeff$^{2}$
\\
$^{1}$Universit\'e de Paris, CNRS, Astroparticule et Cosmologie, F-75013 Paris, France\\
$^{2}$Universit\'e Paris-Saclay, CNRS/IN2P3, IJCLab, 91405 Orsay, France\\
}

\date{Accepted XXX. Received YYY; in original form ZZZ}

\pubyear{2019}

\begin{document}
\label{firstpage}
\pagerange{\pageref{firstpage}--\pageref{lastpage}}
\maketitle

\begin{abstract}
We model the diffusive shock acceleration of particles in a system of two colliding shock waves and present a method to solve the time-dependent problem analytically in the test-particle approximation and high energy limit. In particular, we show that in this limit the problem can be analysed with the help of a self-similar solution.
While a number of recent works predict hard ($E^{-1}$) spectra for the accelerated particles in the stationary limit, or the appearance of spectral breaks, we found instead that the spectrum of accelerated particles in a time-dependent collision follows quite closely the canonical $E^{-2}$ prediction of diffusive shock acceleration at a single shock, except at the highest energy, where a hardening appears, originating a bumpy feature just before the exponential cutoff.
We also investigated the effect of the reacceleration of pre-existing cosmic rays by a system of two shocks, and found that under certain conditions spectral features can appear in the cutoff region.
Finally, the mathematical methods presented here are very general and could be easily applied to a variety of astrophysical situations, including for instance standing shocks in accretion flows, diverging shocks, backward collisions of a slow shock by a faster shock, and wind-wind or shock-wind collisions.
\end{abstract}

\begin{keywords}
acceleration of particles -- shock waves -- cosmic rays
\end{keywords}



\section{Introduction}

Astrophysical shocks are particle accelerators.
Direct evidence for particle acceleration at shocks comes from the exploration of interplanetary space \citep[e.g.][]{reames1999}, while indirect evidence comes from the detection of the radiation produced by energised particles. Radiative signatures of particle acceleration at shocks have been observed from many astrophysical objects, spanning from galactic (e.g. supernova remnants, \citealt{helder2012}) to cosmological (e.g. structure formation shocks, \citealt{vanweeren2010}) scales.
Moreover, it is believed that most of the observed cosmic rays are accelerated at shocks
\citep[e.g.][for a recent review]{gabici2019}.

Colliding shocks are encountered in a great variety of astrophysical contexts, spanning from the interplanetary medium (e.g. shocks in the solar wind, \citealt{colburn1966}), the interstellar medium (e.g. wind-wind or supernova-wind or even, occasionally, supernova-supernova shock collisions in stellar clusters or superbubbles, \citealt{bykov2013,parizot2004,reimer2006}).
Collisions between shocks may also take place inside relativistic jets of gamma ray bursts and active galactic nuclei \citep{kobayashi1997,spada2001} or in outflows in novae \citep{steinberg2020}.

In a seminal paper, \cite{axford1990} concluded that the spectrum of particles accelerated at colliding shocks should follow very closely the standard prediction for diffusive shock acceleration at a single shock, i.e. a power law in particle energy $\propto E^{-2}$.
In more recent times, a number of authors reconsidered the problem of particle acceleration at colliding shocks and claimed that indeed some important differences might appear with respect to the standard $E^{-2}$ scenario. In particular, \citet{bykov2013} developed a semi-analytic and non-linear model to describe the acceleration of particles at a couple of colliding shocks, where the pressure of cosmic rays onto the structure of the shocks was also taken into account. To simplify the problem, \citet{bykov2013} assumed the two shocks to be very close to each other, and solved the steady-state (time independent) transport equation for accelerated particles. 
The resulting spectrum of accelerated particles was found to be very hard, scaling as $\propto E^{-1}$. The model was then used to make predictions on the gamma-ray and neutrino emission from colliding winds in compact stellar clusters \citep{bykov2015} and in bow shock wind nebulae \citep{bykov2019}.

\citet{wang2017,wang2019} developed a Monte Carlo code to study the acceleration of particles at a pair of colliding shocks. They considered both the case of converging shocks \citep{wang2017} and that of a faster shock catching up with a slower one \citep{wang2019}. In both cases, they concluded that spectral features such as breaks may appear in the spectrum of accelerated particles.

%
Finally, \citet{siemieniec2000} considered the case of particle acceleration in converging \textit{flows} of plasma, where a pair of standing shocks may appear. They considered the case of accretion of matter onto cosmological structures, and concluded that the resulting spectrum of accelerated particles is a power law in energy characterised by a very hard slope: $\propto E^{-1}$. This solution is identical to that found by \citet{bykov2013} for the case of acceleration at colliding shocks. This is not surprising as both approaches are based on the steady-state transport equation.

Motivated by these recent studies, in this paper we investigate the acceleration of particles at a pair of colliding shocks. We tackle the problem both numerically and analytically, and demonstrate that under certain conditions an analytic self-similar solution can be found asymptotically. To our knowledge, this is the first attempt to solve analytically the time-dependent problem of particle acceleration at colliding shocks. We found that the spectrum of accelerated particles follows quite closely the canonical $E^{-2}$ prediction of diffusive shock acceleration at a single shock, except at the highest particle energies, where a hardening appears, originating a bumpy feature just before the exponential cutoff. The $E^{-1}$ slope is recovered in our approach only as an asymptotic limit, which is in fact never satisfied due to the finite age of the accelerator (i.e. particles are no longer accelerated when the shocks collide). 
We also investigated the effect of the reacceleration of pre-existing cosmic rays by a system of two shocks, and found that under certain conditions spectral features can appear in the cutoff region.


The paper is structured as follows. In Section \ref{sec:model} we describe the model and discuss the transport equation. Such equation is solved numerically in Section \ref{sec:numerical_resolution} and an analytical analysis is performed in Section \ref{sec:analytic}, where we show how the problem can be handled by means of an hypothesis of self-similarity. In Section \ref{sec:standing} we briefly discuss the case of a pair of standing shocks in a converging flow of matter. We conclude in Section \ref{sec:conclusions}.


\section{Particle acceleration at converging shocks: setup}
\label{sec:model}

\subsection{Model}

\begin{figure}
	\includegraphics[width=\columnwidth]{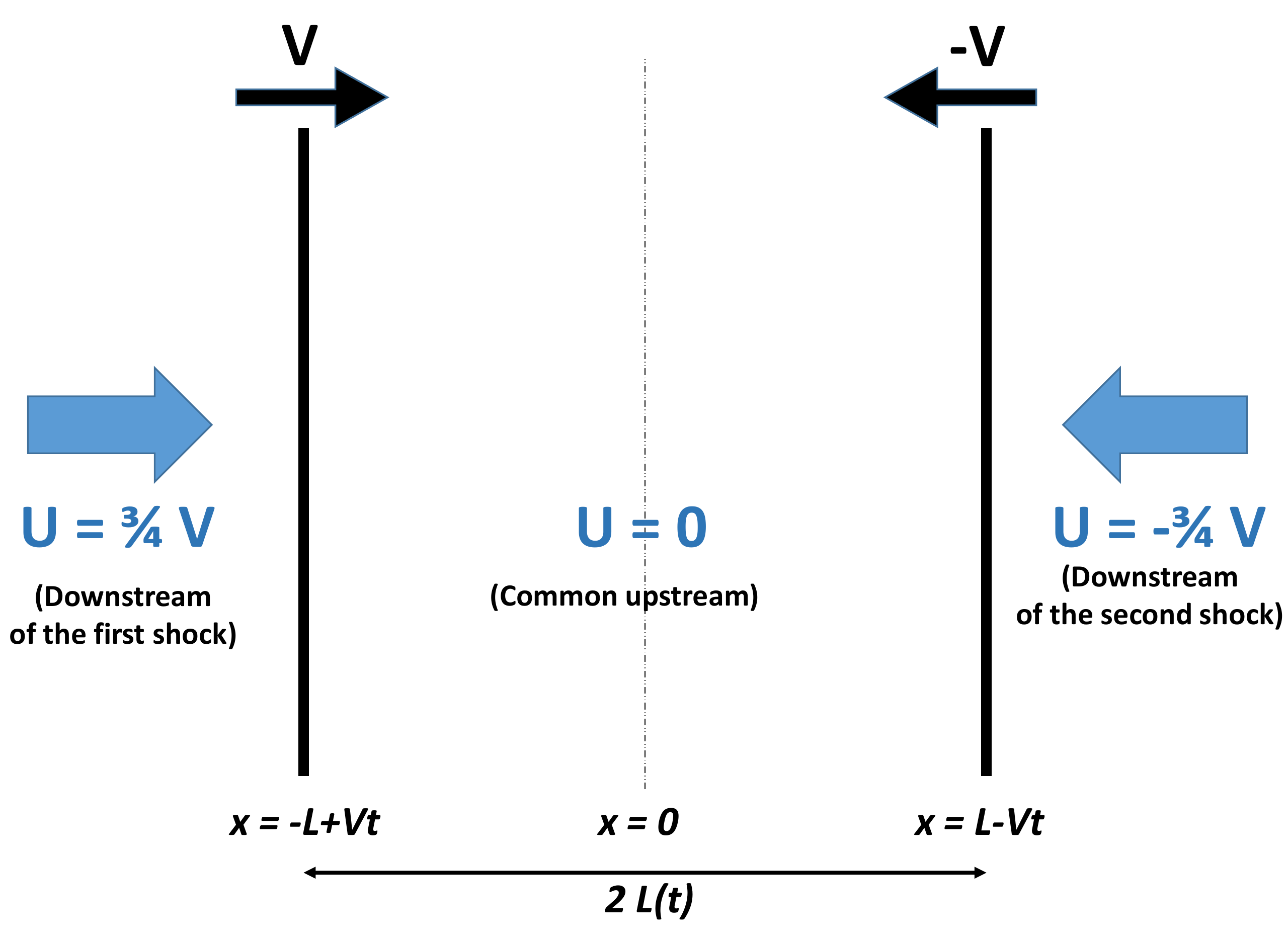}
    \caption{Setup of the problem.}
    \label{fig:generalsystem}
\end{figure}

We consider a system of two infinite and plane shocks initially (time $t = 0$) separated by a distance $2 L$. The two shock surfaces are parallel to each other, and move along the same direction (defined as the $x$ axis) at a velocity $V$ and $-V$, respectively. The origin of the $x$ axis is defined in such a way that the positions of the shocks at a given time $t$ are $x_s = \pm (L - V t)$. In other words, the system is symmetric with respect to $x = 0$, and the shocks will collide at a time $t_{\rm coll} = L/V$.
The initial setup of the problem is shown in Fig.~\ref{fig:generalsystem}, where one can see that while the upstream medium between the shocks is at rest, the downstream medium moves in the same direction of the shock with a speed equal to $3/4 \times V$. Here, we have assumed strong shocks with compression ratio $r=4$.

We now model the acceleration of particles at the system of two shocks in the test-particle limit, i.e., the pressure of accelerated particles does not affect the shock structure. 
%
We also assume a spatially homogeneous and steady diffusion coefficient for accelerated particles, both upstream and downstream of the shock, and neglect energy losses.


Under these simplifying assumptions, the evolution of the particle distribution function $f(x,p,t)$ can be derived from the cosmic ray transport equation \citep[e.g.][]{kirk1994}:
\begin{equation}\label{CRtransport}
\begin{split}
\d_t f(x,p,t) + u \d_x f(x,p,t) = \d_{x} \( \kappa \d_{x} f(x,p,t) \) + \\ \frac{1}{3} (\d_{x} u) p \d_p f(x,p,t) + Q_i \delta(x \pm x_s) \delta(p-p_i) ~,
\end{split}
\end{equation}
where $u$ is the plasma velocity, $\kappa$ the particle diffusion coefficient, and $Q_i$ the injection at the shocks of particles of momentum $p_i$. 
Finally, $\delta(x \pm x_s) = \delta(x + x_s) + \delta(x - x_s)$, where $\pm x_s$ are the positions of the shocks.

We rescale Equation \ref{CRtransport} with the following space dilatation:
\begin{equation}
\label{eq:zoom}
X \equiv x/(L-Vt)
\end{equation}
such that in the new rest frame the two shocks stand at a fixed position $X = \pm 1$. We also define the following dimensionless quantities:
\begin{align}\label{dimensionlessq}
\begin{aligned}
U &\equiv u/V \, , \quad K \equiv \kappa/(LV) \, , \quad
P \equiv p/p_i~, \\
\tilde{t} &\equiv V t/L \, , \quad T \equiv -\ln\(1- \tilde{t}\)~,
\end{aligned}
\end{align}
in order to obtain the following dilated dimensionless transport equation:
\begin{equation}\label{dimensionlessCRtransport}
\begin{split}
\d_T f + (X + U) \d_X f
= \d_X \( e^T K \d_X f \)
+ \frac{\d_X U}{3} P \d_P f + Q~,
\end{split}
\end{equation}
where $Q = \delta(X \pm 1) \delta(P-1)$ (the injection density $Q_i/V/p_i$ has been absorbed in $f$). The additional term $X \d_X$ comes from a fictitious velocity due to the fact that we are gradually zooming in the space coordinate as time flows. The factor $e^{T}$ in front of the diffusion coefficient accounts for the increase of the diffusion length over separation length ratio: as in the new rest frame the separation length remains constant in time, the diffusion length should effectively vary in time.

As we have two length scales, the diffusion length and the separation distance, it is not possible to absorb both by a redefinition of the spatial coordinate, which is why the diffusion coefficient still appears in the dimensionless transport equation \ref{dimensionlessCRtransport}.
However, provided that we assume some shape for the diffusion coefficient $K$ as function of momentum,
\begin{equation}
    K = K_0 \psi(P)~,
\end{equation}
we can translate the time as:
\begin{equation}\label{time_redef}
\Hat{T} \equiv T + \ln K_0 = - \ln\(\frac{1- \tilde{t}}{K_0}\)
\end{equation}
in order to get rid of at least $K_0$ in the transport equation:
\begin{equation}\label{transport_Equation_for_resolution}
\begin{split}
\d_{\Hat{T}} f + (X + U) \d_X f
= \d_X \( e^{\Hat{T}} \psi(P) \d_X f \)
+ \frac{\d_X U}{3} P \d_P f  + Q~,
\end{split}
\end{equation}
at the price of a shifted initial time:
\begin{equation}
    \Hat{T}_0 = \ln K_0 ~.
\end{equation}

As $K$ is the ratio between the diffusion length and the initial distance between the shocks (see Equation \ref{dimensionlessq}),
$K_0$ is a parameter describing how far away are the shocks from each other at the beginning of the process. The parameter $K_0$ will be used for quantitative comparisons between the acceleration of particles at two converging shocks and at a single shock. In particular, the case $K_0 = 0$ corresponds to two shock at an infinite distance that would behave as two isolated shocks. More importantly, the introduction of the parameter $K_0$ also shows that the result of Eq.~\ref{transport_Equation_for_resolution} will only depend on the normalisation of the diffusion coefficient through the ratio between diffusion length and initial shock distance.

\subsection{Timescales}
\label{sec:timescales}

The characteristic diffusion length of particles of energy $E$ at a single shock is given by the well known expression \citep[e.g.][]{drury1991}:
\begin{equation}
l_d(E) = \frac{\kappa(E)}{V} ~.
\end{equation}
Particles of energy $E$ fill a region of size $l_d$ upstream of the shock. 
Therefore, when the distance between the shocks $2(L-V t)$ equals twice the diffusion distance $2 l_d(E)$, particles of energy $E$ start to be affected by both shocks. 
This happens at a critical time:
\begin{equation}
\label{eq:tc}
    t_c(E) = t_{\rm coll} - \kappa(E)/V^2 ~,
\end{equation}
where $t_{\rm coll} = L/V$ is the time at which the collision occurs.

The acceleration time of particles of energy $E$ at a single shock is \citep[e.g.][]{drury1991}:
\begin{equation}
\label{eq:acc1}
    t_{\rm acc} = \xi \kappa(E)/V^2 ~,
\end{equation}
where $\xi$ is a dimensionless parameter to be determined later, depending on model assumptions. For a single shock, after a time $t_{\rm acc}(E)$, the spectrum of accelerated particles is a power law up to particle energy $E$.

At this point, we can define an energy-dependent dimensionless parameter $r_{\rm eq}$ as:
\begin{equation}
\label{eq:r_eq}
    r_{\rm eq}(E) \equiv \frac{t_c}{t_{\rm acc}} = \frac{1}{\xi} \(\frac{LV}{\kappa(E)} - 1\) = \frac{1}{\xi} \(\frac{1}{K} - 1\) ~.
\end{equation}
The case $r_{\rm eq}(E) \ll 1$ corresponds to $t_c \ll t_{\rm acc}$, i.e., to a situation where the two shocks become very close to each other before they had time to accelerate particles up to an energy $E$.
On the other hand, $r_{\rm eq}(E) \gg 1$ implies that particles can be accelerated at a single shock up to an energy $E$ well before both of the shocks begin to contribute to the acceleration of particles of that energy.

The case $r_{\rm eq}(E_{\rm eq}) = 1$ defines a characteristic energy $E_{\rm eq}$ and gives:
\begin{equation}
    \kappa (E_{\rm eq}) = \frac{LV}{\xi + 1}
\end{equation}
or, using Eq.~\ref{eq:tc}:
\begin{equation}
    t_c^{\rm eq} = \frac{\xi}{1+\xi} t_{\rm coll} ~.
\end{equation}
The physical meaning of $t_c^{\rm eq}$ is the following: for times earlier than $t_c^{\rm eq}$ the two shocks accelerate particles as two isolated systems, while after $t_c^{\rm eq}$ the two shocks behave as a single particle accelerator.

Let's now determine the value of the parameter $\xi$.
The acceleration time at a single shock is defined as \citep{drury1991}:
\begin{equation}
    t_{\rm acc} = \frac{3}{u_1-u_2} \int_{p_0}^p \frac{\dd p'}{p'} \( \frac{\kappa_1(p')}{u_1} + \frac{\kappa_2(p')}{u_2} \)~.
\end{equation}
For strong shocks $u_2 = u_1/4 = V/4$. We further assume that $\kappa_1$ and $\kappa_2$ are spatially uniform and scale as in Bohm diffusion ($\kappa_{1,2} \propto p$). Moreover, we set $\kappa_2 = \alpha \times \kappa_1$. It is generally believed that, due to magnetic field compression and generation of magnetic turbulence downstream of the shock $\alpha < 1$. 
Then, one can easily see that $\xi = 4 (1+4\alpha)$ and that for $0 < \alpha < 1$ one gets:
\begin{equation}
t_c^{\rm eq} = 0.8 ... 0.95 \times t_{\rm coll}
\end{equation}
meaning that only during the late phase of the collision process accelerated particles are affected by both shocks simultaneously.
This also implies that, as  pointed out by \citet{axford1990}, the maximum particle energy attainable at such systems will most likely not change dramatically with respect to that expected at a single shock.
For example, assuming Bohm diffusion ($\kappa = \kappa_0 E$) the maximum energy will be $E_{\rm max} \gtrsim E_{\rm eq} \sim  LV/(1+\xi)/\kappa_0$, 
which is similar to the maximum energy obtained by equating $t_{\rm acc}$ for a single shock (Eq.~\ref{eq:acc1}) with $t_{\rm coll} = L/V$: $E_{\rm max} = LV/(\xi \kappa_0)$. 

Therefore, as we will confirm in the following, the spectrum resulting from the acceleration of particles at a pair of converging shocks will not change significantly as long as low energies (i.e. much smaller than $E_{\rm eq}$) are considered. However, the shape of the spectral cutoff could be affected. 
Therefore, the study presented here can have a significant impact on the interpretation of the cutoffs observed in the non-thermal spectra of astrophysical sources. The study of spectral cutoffs is of paramount importance in order to constrain the physical properties of astrophysical accelerators \citep{romoli2017}.

\section{Numerical solution}
\label{sec:numerical_resolution}

In this Section we present a numerical solution of Eq.~\ref{transport_Equation_for_resolution}, which provides the spatial distribution and the spectrum of accelerated particles at the system of two converging shocks.

\subsection{Methods}
Since the system of approaching shocks considered here is symmetric with respect to $x = 0$, we only solve the transport equation for the half space $x<0$, imposing that the derivative of the particle density distribution vanishes at $x=0$ and that the function itself vanishes at $x \to -\infty$. In order to account for the whole space from $-\infty$ to $0$, we further change variable in Equation \ref{transport_Equation_for_resolution}, defining $Z \equiv \exp \(X+1\)$ such that $Z$ goes from $0$ to $e$, with the shock at $Z=1$.  We finally change the momentum variable for $Y = \ln P$. The transformed transport equation reads:
\begin{equation}
\label{eq:solve}
\begin{split}
\d_{\hat{T}} f + W Z \d_Z f
= \tilde{K} Z^2 \d^2_Z f + \delta(Z-1) \( \frac{\Delta U}{3} \d_Y f + \delta(Y) \)~,
\end{split}
\end{equation}
where $\tilde{K} = \psi(P) e^{\hat{T}}$, $W = (\ln Z - 1 + U -\psi)$ and $\Delta U = U_2-U_1$.

Our numerical scheme follows that presented in \citet{drury1991}: the spatial variable is treated using a Crank-Nicholson method, while the momentum transfer at the shock is updated at each time step, starting from the injection momentum $P=1$.

In order to define the problem, the only parameter to be specified is the shifted initial time $\hat{T}_0 = \ln{K_0}$. It should be stressed that a smaller value of $K_0$ translates into an higher maximum energy. This can be understood in two different equivalent ways: {\it i)} if the initial time is smaller there is more time to accelerate particles before the shock collision happens; {\it ii)} Bohm diffusion implies $r_{\rm eq} \propto 1/K_0$, so that a small $K_0$ results in $r_{\rm eq} \gg 1$ and the considerations made in Sec.~\ref{sec:timescales} apply. However, as $L \propto 1/K_0$, the smaller $K_0$ the bigger the initial physical distance between the shocks, hence one needs a finer spatial resolution to solve the problem with small values of $K_0$, which increases the computation time. Eventually we found $K_0 = 0.003$ to be a satisfying compromise between the required accuracy and computation time. 

\subsection{Results}
\label{sec:numsolresults}
\begin{figure}
	\includegraphics[width=\columnwidth]{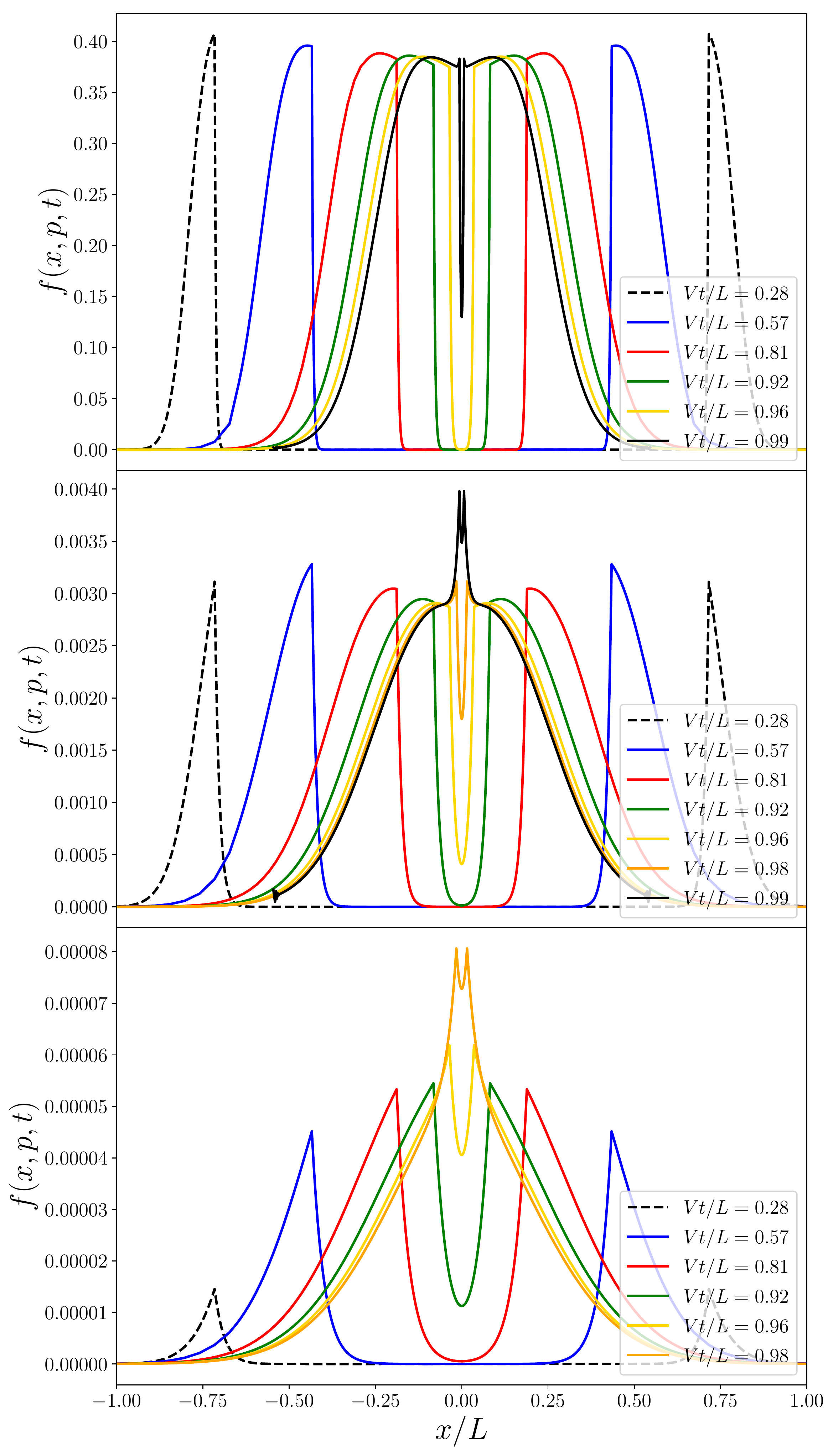}
    \caption{Spatial distribution of accelerated particles (arbitrary dimensionless units) for three different particle momenta, as given by the dimensionless parameter $r_{\rm eq}$ (Eq.~\ref{eq:r_eq}). From top to bottom: $r_{\rm eq} = 13$, $r_{\rm eq} =3.7$, $r_{\rm eq} = 1.3$.}
    \label{fig:shapes_converging}
\end{figure}

Figure \ref{fig:shapes_converging} shows the solution of Eq.~\ref{eq:solve}. The spatial distribution of accelerated particles is plotted for three different particle momenta, corresponding to $r_{\rm eq} \gg 1$ (top panel), $r_{\rm eq} \gtrsim 1$ (middle panel), and $r_{\rm eq} \sim 1$ (bottom panel).
The sharp features in the curves indicate the position of the shock waves. 

At small momenta ($r_{\rm eq} \gg 1$, top panel) and small times, one recovers the solution for particle acceleration at a single shock, characterised by an exponential decay of $f$ upstream of the shock. Then, as time increases, the gradient of the particle distribution ahead of the shock gradually decreases because of the influence of the other shock.  
When the two shocks are very close, a sharp peak at the shock position also appears.
This happens when the value of $f$ at $x = 0$ is no longer negligible, i.e., when the number of particles coming from the other shock exceeds the number of particles injected there.

The origin of the bump visible downstream of the shocks at late times (top panel in Fig.~\ref{fig:shapes_converging}) can be better understood by rewriting the transport equation (Eq.~\ref{transport_Equation_for_resolution}) in terms of the time variable $\tau \equiv e^{\hat{T}} - 1$:
\begin{equation}
\begin{split}
\d_{\tau} f + \frac{1}{1+\tau}(X+U) \d_X f
= \d_X \( \psi \d_X f \) + \\
\frac{1}{1+\tau} \left[ \frac{\d_X U}{3} P \d_P f + \delta(X \pm 1) \delta(P-1) \right] ~,
\end{split}
\end{equation}
which is a diffusion (first term on the right) advection (second term on the left) equation, with an effective injection represented by the second term on the right.
As time passes, advection and injection at all momenta are less and less efficient, as the corresponding terms in the Equation scale as $1/(1+\tau)$.
The reduced injection induces a suppression in the value of $f$ at the shock, and the reduced advection makes particles less capable to reach large distances in the downstream region, and this creates a bump. 

For high momenta ($r_{\rm eq} \sim 1$, bottom panel in Fig.~\ref{fig:shapes_converging}), the shocks become very close to each other before being able to accelerate a significant number of particles, and this explains the increase in time of the maximum of $f$.
The double-peak feature visible at very late times has the same origin as in the top panel. The intermediate case $r_{\rm eq} \gtrsim 1$ is shown in the middle panel.


\begin{figure}
	\includegraphics[width=\columnwidth]{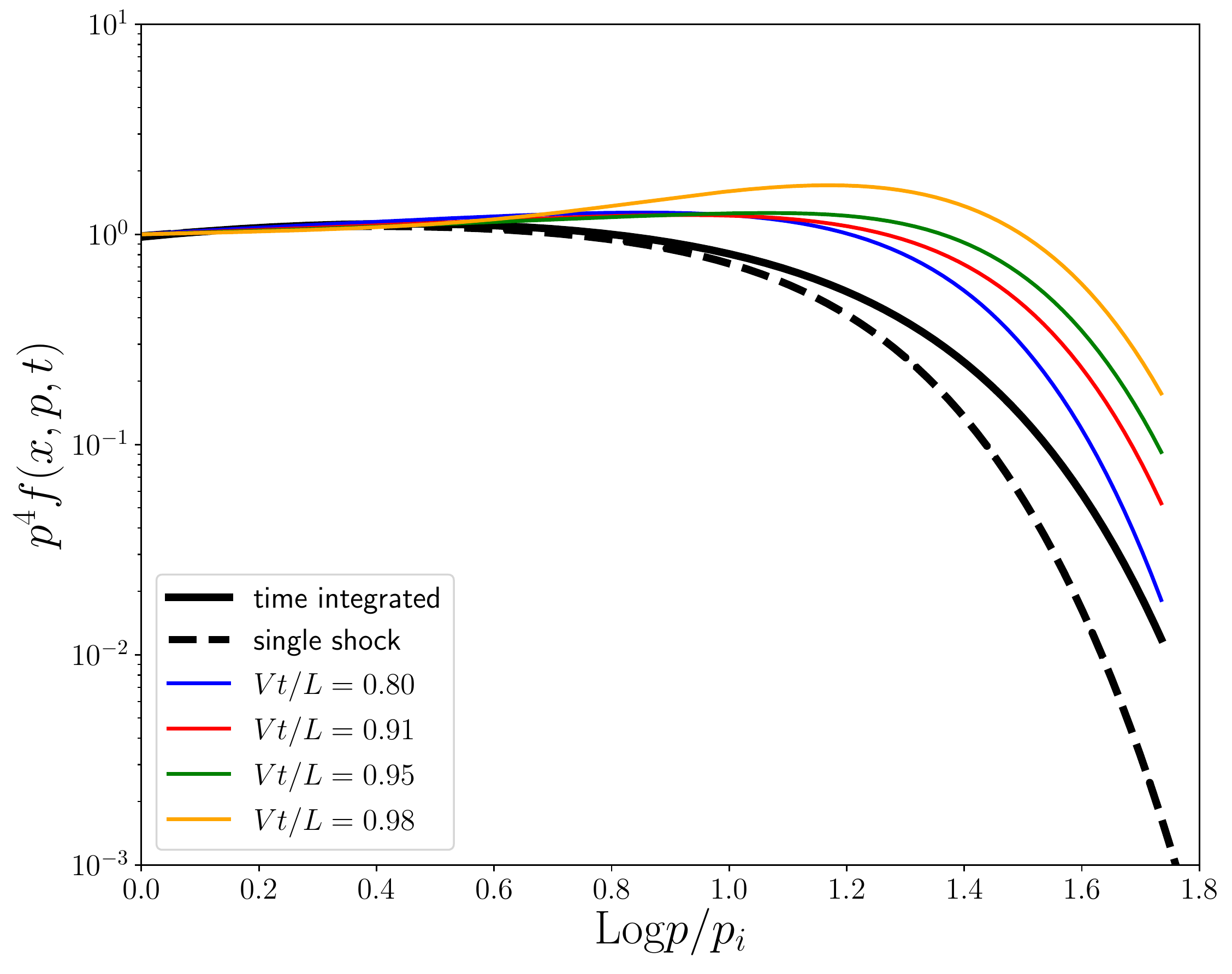}
    \caption{Spectra of accelerated particles at the shock location are shown with coloured thin lines for different times. 
    The black solid line represents the spectrum of particles integrated over the entire volume at the time of shock collision (or, equivalently, the total spectrum of particles accelerated during the entire lifetime of the system). This is compared with the same quantity one would obtain for a single shock (dashed line).}
    \label{fig:spectra_converging}
\end{figure}

Figure \ref{fig:spectra_converging} shows the particle spectra obtained at the shock position for different times, compared to the solution for a single shock.
For small times and momenta, one recovers the canonical spectral index close to $-4$. However, at energies such that the diffusion length equals half the distance between the shocks, each shock begins to be fed by the other, such that there are more energetic particles around it: this is why, at times close to the collision, one sees a bump in the spectrum at high energies. 
The bump is averaged out when computing the time-integrated spectrum over the entire lifetime of the system (black solid line in Fig.~\ref{fig:spectra_converging}) but the shape of the cut-off is broader than what expected for particle acceleration at a single shock (black dashed line in the Figure). 

Apart from this shape modification at the high energy end of the time-integrated spectrum, we did not find any break in the spectrum, which remains very close to the canonical $p^{-4}$ power law for all particle energies significantly smaller than the cutoff. 


\section{Analytic approach}
\label{sec:analytic}

In this Section we develop an approximate, analytic and self-similar model to describe the acceleration of particles at converging shocks. 
Here, we limit our analysis to the case of Bohm diffusion, $\kappa \propto p$, and we provide the solution for a generic scaling of the diffusion coefficient $\kappa \propto p^{\delta}$ in Appendix~\ref{app1}.

\subsection{The self-similarity hypothesis}
\label{sec:selfsimilartest}

Let us consider particle energies that satisfy the condition $r_{\rm eq}(E) > 1$, where the parameter $r_{\rm eq}$, which was introduced in Sec.~\ref{sec:timescales}, takes values larger than unity when particles of energy $E$ are accelerated at each one of the converging shocks before being affected by the presence of the other shock. In other words, particles reach an energy $E$ at each shock after an acceleration time $t_{\rm acc}(E)$ shorter than the time $t_c(E)$ when they begin to be affected by the presence of the other shock.

\begin{figure}
\centering
  \includegraphics[width=0.5\linewidth]{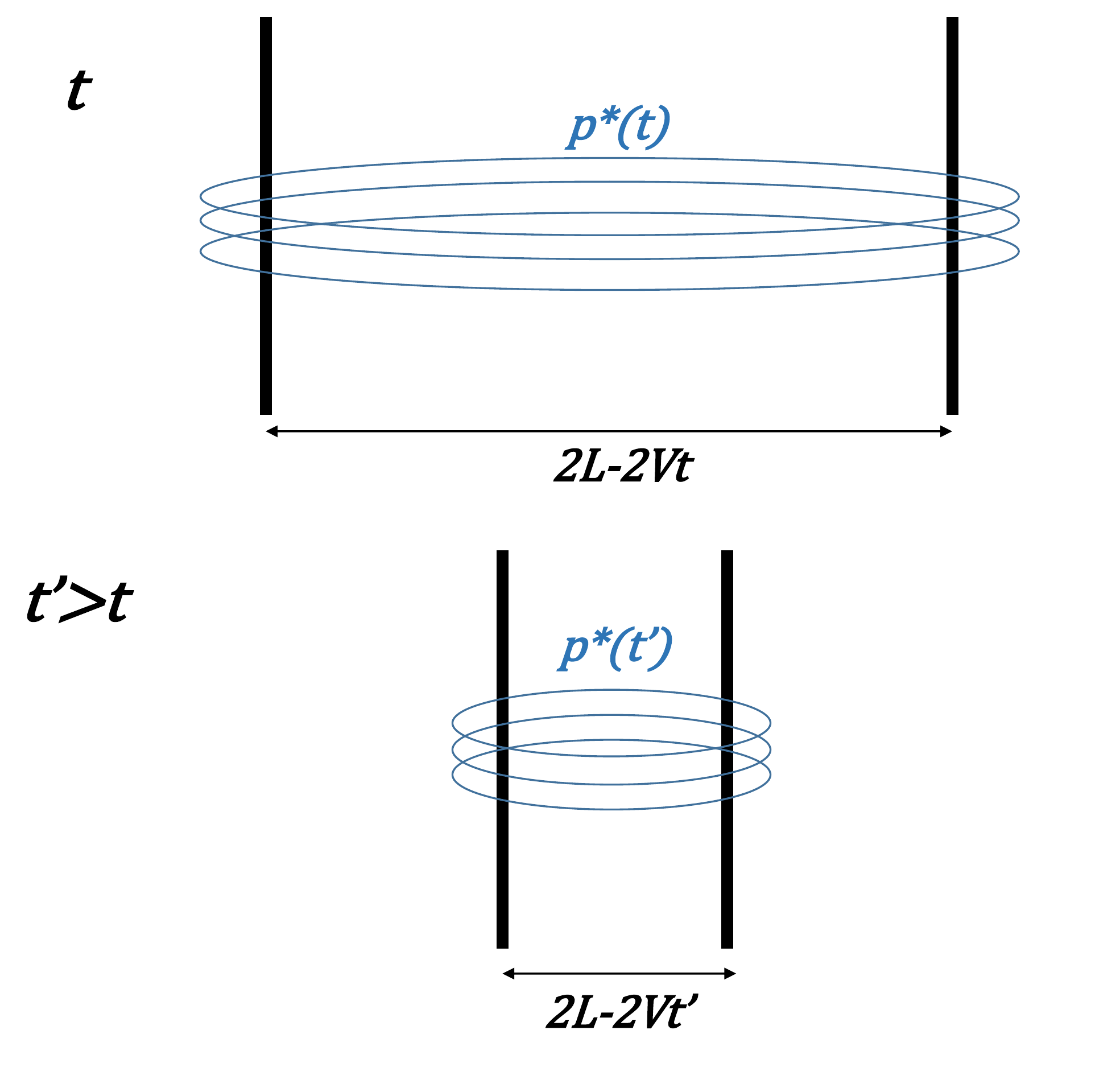}
\caption{\label{fig:rescaling}
The cartoon shows the two shocks at different times. The thin blue lines represent the diffusion length of particles of a given momentum. 
One can see that particles of a given momentum $p^*(t)$ at a given time $t$ behave like particles of lower momentum $p^*(t^{\prime})$ at a later time $t^{\prime}$.}
\end{figure}

The top panel of the cartoon in Fig.~\ref{fig:rescaling} represents the converging shock system at a given time $t$. By imposing $t_c(E) \sim t$ one can derive an energy $E_*(t)$ (or momentum $p_*(t)$) such that particles of that energy begin at a time $t$ to be affected by both shocks.
The bottom panel of the Figure shows the system at a later time $t^{\prime} > t$. This Figure illustrates that particles of momentum $p_*(t^{\prime})$ will be, at time $t^{\prime}$, in the very same situation as particles of larger momentum $p_*(t)$ at the earlier time $t$.
The two momenta are connected by the relation:
\begin{equation}
\label{eq:pscale}
p_*(t^{\prime}) = \frac{L-V t^{\prime}}{L-V t} p_*(t)~,
\end{equation}
where we have assumed again Bohm diffusion. The more general case $\kappa \propto E^{\delta}$ would give $p_*(t^{\prime}) = [(L-V t^{\prime})/(L-V t)]^{1/\delta} p_*(t)$. 

The relation between momenta $p_*$ at different times (Eq.~\ref{eq:pscale}) is identical to that connecting the coordinate $x$ of a point as seen from the new reference frame introduced by Eq.~\ref{eq:zoom}. Indeed, for a fixed value of the new spatial coordinate $X$ the scaling is:
\begin{equation}
\label{eq:xscale}
x(t^{\prime}) = \frac{L-V t^{\prime}}{L-V t} x(t) ~ .
\end{equation}

This motivated us to search for a self-similar (scale-free) solution of the acceleration problem, that mathematically translates into the requirement:
\begin{equation}\label{scale_invariant_f}
f(x,p,t) = \lambda^{\alpha} f(\lambda x, \lambda p, t_0)
\end{equation}
with
\begin{equation}
\label{eq:lambda}
\lambda = \frac{L-V t_0}{L-V t}
\end{equation}
and with the additional requirement (see discussion at the beginning of this Section):
\begin{equation}
r_{\rm eq}\( \lambda p \) > 1 ~.
\end{equation}

In order to determine the value of the parameter $\alpha$ it is convenient to compute the time derivative of $f(x,p,t)$ as given by Eq.~\ref{scale_invariant_f}:
\begin{equation}\label{time_derivative_1}
\d_t f(x,p,t) = \frac{V}{L-V t} \( \alpha  f\(x,p,t\) + x \d_x f\(x,p,t\) + p \d_{p} f\(x,p,t\) \) ~. 
\end{equation}
We know that at any time, for any $p<p_*(t)$, i.e. for all particles that are not able to see both shocks, the solution should be the one obtained in the case of a single shock acceleration, which is time invariant and spatially constant in a region downstream of the shock of length $\ll Vt/4$.
Under these circumstances Eq.~\ref{time_derivative_1} reduces to:
\begin{gather}\label{time_derivative_2}
\alpha f\(x,p,t\) + p \d_{p} f\(x,p,t\) = 0 ~, 
\end{gather}
which shows that in order to recover the solution $f \sim p^{-4}$ for particles accelerated at a single shock the value of the parameter $\alpha$ must be equal to 4.

The numerical approach described in Section \ref{sec:numerical_resolution} allows to test the hypothesis of self-similarity.
\begin{figure*}
\centering
  \includegraphics[width=1\linewidth]{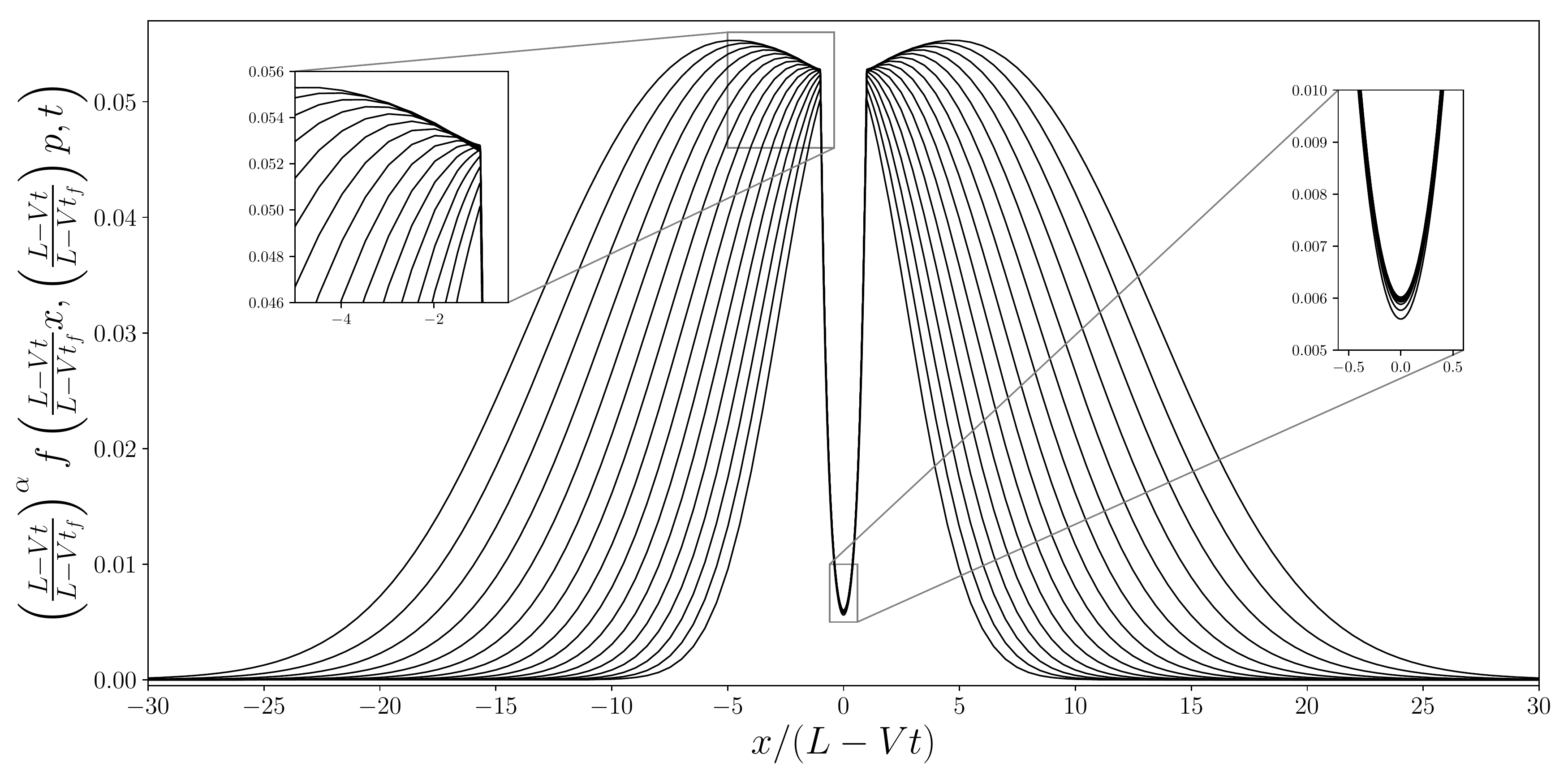}
\caption{\label{fig:shape_rescaled}Spatial distribution of accelerated particles in rescaled coordinates (see text). Curves refer to different times between $Vt_i/L = 0.9$ and $Vt_f/L = 0.98$, and for a momentum at $t_i$ equal to $p/p_i = 10$. The best match between solutions at different times is found for values of $\alpha$ gradually approaching $\sim$4 as we increase the resolution of the numerical scheme.}
\end{figure*}
Figure~\ref{fig:shape_rescaled} shows the results of our numerical computation for the particle distribution function $f$ plotted at different times, with particle momenta and space coordinates rescaled as in Eqs.~\ref{eq:pscale} and \ref{eq:xscale}, and multiplied by $\lambda^{\alpha}$ (see Eq.~\ref{eq:lambda}). For the highest resolution numerical calculation that we performed we found that the best match is obtained by setting $\alpha = 3.9$. The small discrepancy with respect to the analytic prediction ($\alpha = 4$) is most likely due to the numerical accuracy of our approach. We performed the same numerical calculation for different resolutions: the best value for $\alpha$ gets closer to 4 as the numerical resolution increases.


\begin{figure}
\centering
  \includegraphics[width=1\linewidth]{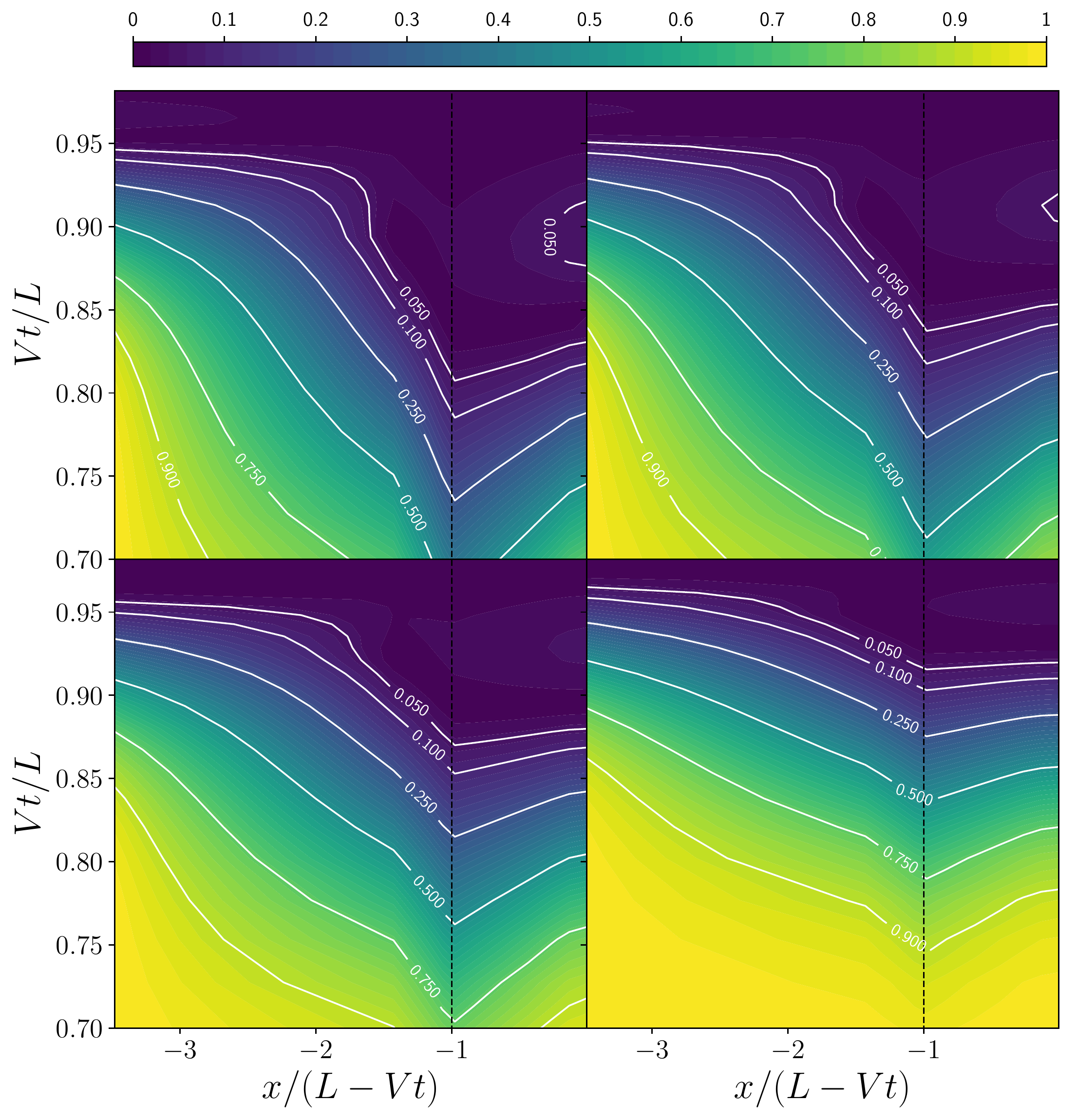}
\caption{The color scale refers to the value of the quantity defined in Eq.~\ref{test_selfsimi}, i.e. the discrepancy between the self-similar solution and the numerical one at a time $t_f = 0.98$ (see text for more details). The self-similar solution is computed rescaling the solution at an earlier normalised time $Vt/L$.
The four panels (from top left to bottom right) refer to momenta equal to $p/p_i =$~16, 19, 25, 40 at the normalised time $Vt/L = 0.7$.}
\label{fig:scaleinv_maps}
\end{figure}

To be more quantitative on this point, we plot in Figure~\ref{fig:scaleinv_maps} the distribution in the plane $(x,t)$ of the following quantity:
\begin{equation}\label{test_selfsimi}
    \left| \lambda^\alpha f \(\lambda x,\lambda p,t \) - f \(x,p,t_f \) \right| / f \(x,p,t_f \) \, , 
\end{equation}
at four different momenta $p$, where $\lambda= (L-Vt)/(L-Vt_f)$.
The quantity defined in Eq.~\ref{test_selfsimi} measures the error between the distribution function obtained at time $t_f$ after rescaling the solution at earlier time $t$ by means of Eq.~\ref{scale_invariant_f}, and the one directly computed numerically at time $t_f$. 
We show the result for $t_f = 0.98$, i.e. very close to the collision ($t_f = 1$). Fig.~\ref{fig:scaleinv_maps} shows that the self-similarity hypothesis holds well upstream of the shock, as well as in a region downstream of it. The agreement between the self-similar and the numerical solution is good whenever the energy-dependent criterion $r_{\rm eq} > 1$ is fulfilled (blue regions in the map).


\subsection{The self-similar transport equation}


At this point, we can substitute the scaled solution for the particle distribution function $f$ (Eq.~\ref{scale_invariant_f}, with $\alpha = 4$) into the transport equation (Eq.~\ref{CRtransport}) to obtain the stationary differential equation:
\begin{equation}\label{selfsimilar_transport_Equation}
\begin{split}
4 f + (X + U) \d_X f - \tilde{K} \d^2_{X} f = \\ \( \frac{\d_{X} U}{3}-1 \) P \d_P f + \tilde{Q} \delta(X \pm 1) \delta (P-1) ~,
\end{split}
\end{equation}
where $X = x/(L-Vt)$, $P = p/p_i$, $\tilde{K} = \frac{K(P) L}{L-Vt}$, $\tilde{Q} = \( \frac{L-Vt}{L-V t_0} \)^4 \frac{Q}{V}$.
Note that the only physical parameter is the normalised diffusion coefficient $\tilde{K}$, so that the high energy limit is equivalent to $\tilde{t} \to 1$. This reflects the idea of self-similarity: at times close to the collision, particles of any energies cross both shocks, as was the case earlier for particles of high energy only.

\subsubsection{Spectral hardening at high energies}
\label{sec:hardening}

Assuming a power-law for the distribution function $f \sim P^{-4 \beta}$, Eq.~\ref{selfsimilar_transport_Equation} reduces upstream (in between the shocks, $U=0$) to:
\begin{equation}
\label{step1}
4 (1-\beta)f + X \d_X f - \tilde{K} \d^2_{X} f = 0~.
\end{equation}
The solution at the location of one shock ($X \to -1^+$) is expressed in terms of hypergeometric functions. Interestingly, these hypergeometric functions are divergent at small momenta or small times ($\tilde{K} \ll 1$) unless $\beta = 1$. We therefore retrieve the standard $P^{-4}$ spectrum in this limit, as expected.

At a generic momentum, evaluating Eq.~\ref{step1} in $X=0$ gives:
\begin{equation}
    \beta  = 1 - \frac{\left. \d_X^2 f \right|_0}{f_0}  \frac{\tilde{K}}{4} < 1 ~.
\end{equation}
The spectrum is therefore expected to be harder at high momenta (or equivalently late times), because both shocks contribute to the acceleration process.
This is in agreement with the results of the numerical solution (see Figure \ref{fig:spectra_converging}): at high momenta, before the cut-off, the spectrum hardens.

In order to derive a precise value for $\beta$, one needs to find $\left. \d_X^2 f \right|_0$, i.e. to solve Eq.~\ref{step1}. This is done in the high energy limit $\tilde{K} \gg 1$ in Appendix \ref{app2}. We found that $\beta = 3/4$, hence one gets a hard spectrum $f \sim p^{-3}$, similarly to what was found by \citet{bykov2013} in its approximate (steady-state) description of converging shocks. 
In fact, a $p^{-3}$ spectrum is never recovered in our numerical results due to the finite lifetime of the system. As a result, a cut-off appears in the spectrum before the asymptotic solution $f \sim p^{-3}$ can be established. This will be investigated further in Section~\ref{sec:spectrum}. 

The result obtained above can even be generalised to non symmetric situations (flows with different velocities). Indeed, it is always possible to move into the reference frame where the shock velocities are opposite (such that the self-similarity hypothesis can be used), at the price of a non zero plasma velocity upstream. However, one can show using a reasoning similar to that of Appendix \ref{app2} that this plasma velocity does not change the result in the limit of $\tilde{K} \gg 1$. This is somehow expected physically: in this regime, particles are not much affected by the upstream flow.
Interestingly, a backward collision between two shocks (i.e. two shocks moving in the same direction with different speeds) have recently been investigated using a Monte-Carlo simulation by \citet{wang2019}, and a hardening of the CR spectrum was reported, in agreement with our findings.
Conversely, our conclusions differ from an earlier claim from the same authors \citep{wang2017} that a collision between two shock fronts may induce a break (steepening) in the spectrum of accelerated particles.
This difference is due, maybe, to the peculiar setup considered in \citet{wang2017}.


\subsection{Time integrated spectrum of cosmic rays accelerated at converging shocks}
\label{sec:spectrum}

In this Section we seek an analytic expression of the time-integrated spectrum of CRs accelerated at converging shocks.

\subsubsection{Phenomenological modelling}
We propose the following approximate expression to describe the spectrum of accelerated particles at the shock position:
\begin{equation}\label{instantaneous_spectrum}
    f = \mathcal{A} P^{-4} \( 1 + \frac{P}{P_c(t)} \) \exp \( \frac{-P}{P_m(t)} \)~,
\end{equation}
where $P_c(t)= (\lambda K_0)^{-1}$ is the particle momentum such that $\tilde{K}(P_c) = 1$ (for particles of momentum $P_c$ the diffusion length is equal to the distance between the two shocks).
The expression above captures all of the features of the solution: the $p^{-4}$ spectrum at low energies, the hardening to $p^{-3}$ at high energies, and the cutoff at a (normalised) particle momentum $P_m(t)$. 
We have seen that the maximum energy that a particle can achieve differs very little if one considers an isolated shock, or a system of two converging shocks. For this reason, we set:
\begin{equation}
    P_m(t) \sim 1 + \frac{V}{\xi K_0 L} t~,
\end{equation}
which corresponds to the maximum particle momentum for acceleration at an isolated shock (see Appendix~\ref{app3} for more details).

\begin{figure}
\centering
  \includegraphics[width=1\linewidth]{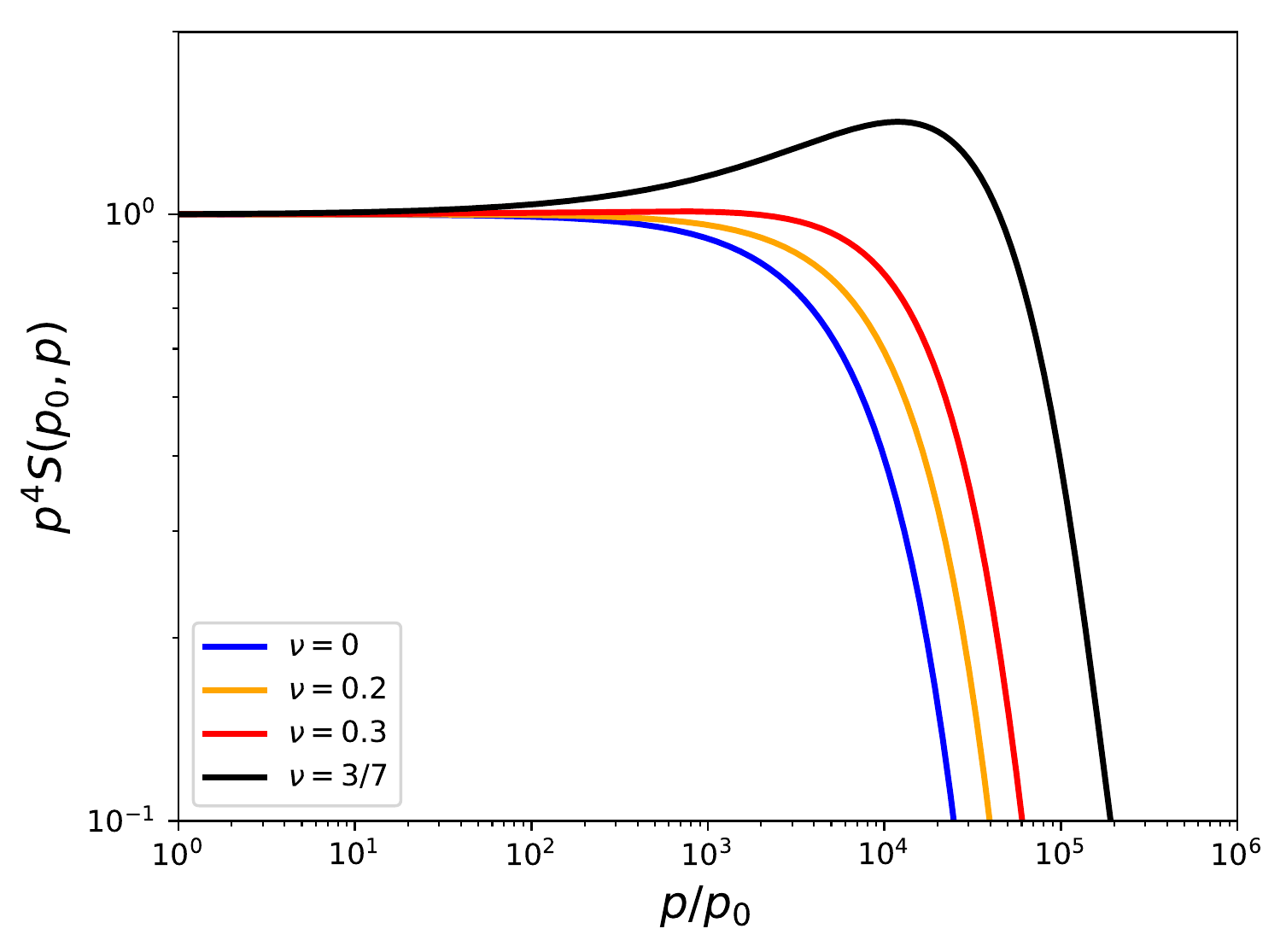}
\caption{\label{fig:green_function}Time-integrated spectra of accelerated particles for $p_0 = 1$ GeV with $V(t) \sim t^{-\nu}$, for various values of $\nu$, taking into account magnetic field amplification and assuming Bohm diffusion. $K_0 = 10^{-6}$ ($\kappa_0 \sim 10^{20}$~cm$^2$/s at $p_0 = 1$ GeV \citep{bykov2013}), $\xi = 20$.}
\end{figure}

We now integrate the instantaneous spectrum (Equation~\ref{instantaneous_spectrum}) over the entire time history of the accelerator, performing the following change of variable:
\begin{equation}
    w \equiv 1 - \lambda^{-1}
\end{equation}
in order to obtain our final result:
\begin{multline}
\label{greenfunctionfinal0}
     S(p_0,p) =  \frac{\mathcal{A}'}{p_0} \( \frac{p}{p_0} \)^{-4} \int_0^1 \dd w  \( 1 + \frac{K(p)}{1-w} \) \exp \( \frac{-K(p)}{K(p_0) + \frac{w}{\xi}} \)~,
\end{multline}
where $K(p)=K_0 p/(m_p c)$ and $\mathcal{A}'$ is a normalisation constant. Plotting the function described in Equation \ref{greenfunctionfinal0} from a low initial momentum $p_0 \sim 1$~GeV, we retrieve a $p^{-4}$ spectrum without any noticeable feature, in agreement with the numerical solution obtained in Section \ref{sec:numsolresults}. This is because the ratio $P_m/P_c$, which drives the appearance of unusual features at the end of the spectrum, increases too slowly. Otherwise stated, particles begin to be affected by both shocks only at the very end of the process. The spectrum for the case of two shocks moving at a constant speed is represented as a blue line in Fig.~\ref{fig:green_function}.

In fact, in a more realistic situation, the colliding shocks might not move at a constant speed, but rather decelerate (e.g. supernova remnant shocks would behave in that way). This is an interesting situation to be investigated because in this case particles would feel the effect of both shocks for a longer time.
We show in Appendix~\ref{app4} how a shock velocity of the form $V(t) \sim t^{-\nu}$ could still be accounted for in our self-similar approach.
The yellow, red, and black line in Fig.~\ref{fig:green_function} refer to the case $\nu =$ 0.2, 0.3, and 3/7 (the latter being the scaling for a Type~I supernova remnant in the free expansion phase according to \citealt{chevalier1982}). It is evident that the position of the spectral cutoff shifts to larger energies when $\nu$ increases, and above a certain value a small bump appears in the spectrum. 

\subsubsection{Spectrum of cosmic rays reaccelerated by converging supernovae shocks}

Time-dependent colliding shocks do not provide strikingly unusual features on cosmic ray spectra, because particles do not have time to be accelerated up to the relevant energies before the end of the process. However, a more significant effect can be expected in the case of the reacceleration of preexisting cosmic rays by colliding shocks, because then particles of high energies already exist at the beginning of the process.


In order to see how the system of converging shocks reaccelerates particles, we assume that at the time of the creation of the converging shocks, preexisting cosmic rays have a standard $p^{-4}$ spectrum with a maximum energy higher than the energy that could be reached by acceleration at the pair of colliding shocks (e.g. cosmic rays confined in a superbubble whose strong turbulence and intermittent supernovae explosions keeps reaccelerating them up to PeV energies \citealt{ferrand2010}). Results can be easily generalised to the case of a spectrum of preexisting particles different than $p^{-4}$.

The spectrum after reacceleration is computed according to:
\begin{equation}
    f(p) = \int_{p_i}^p \dd p_0 \(\frac{p_0}{p_i} \)^{-4} S(p_0,p)~,
\end{equation}
using Equation \ref{greenfunctionfinal0} for $S(p_0,p)$, where the normalization constant is adjusted numerically to impose the conservation of particles (i.e. $\int \dd p S(p_0,p) = 1$). 
We set the injection $p_i = 10$~MeV and $K_0 = 10^{-6}$. The parameter $\xi$ is set to 20 as expected assuming a homogeneous diffusion coefficient.

\begin{figure}
\centering
  \includegraphics[width=1\linewidth]{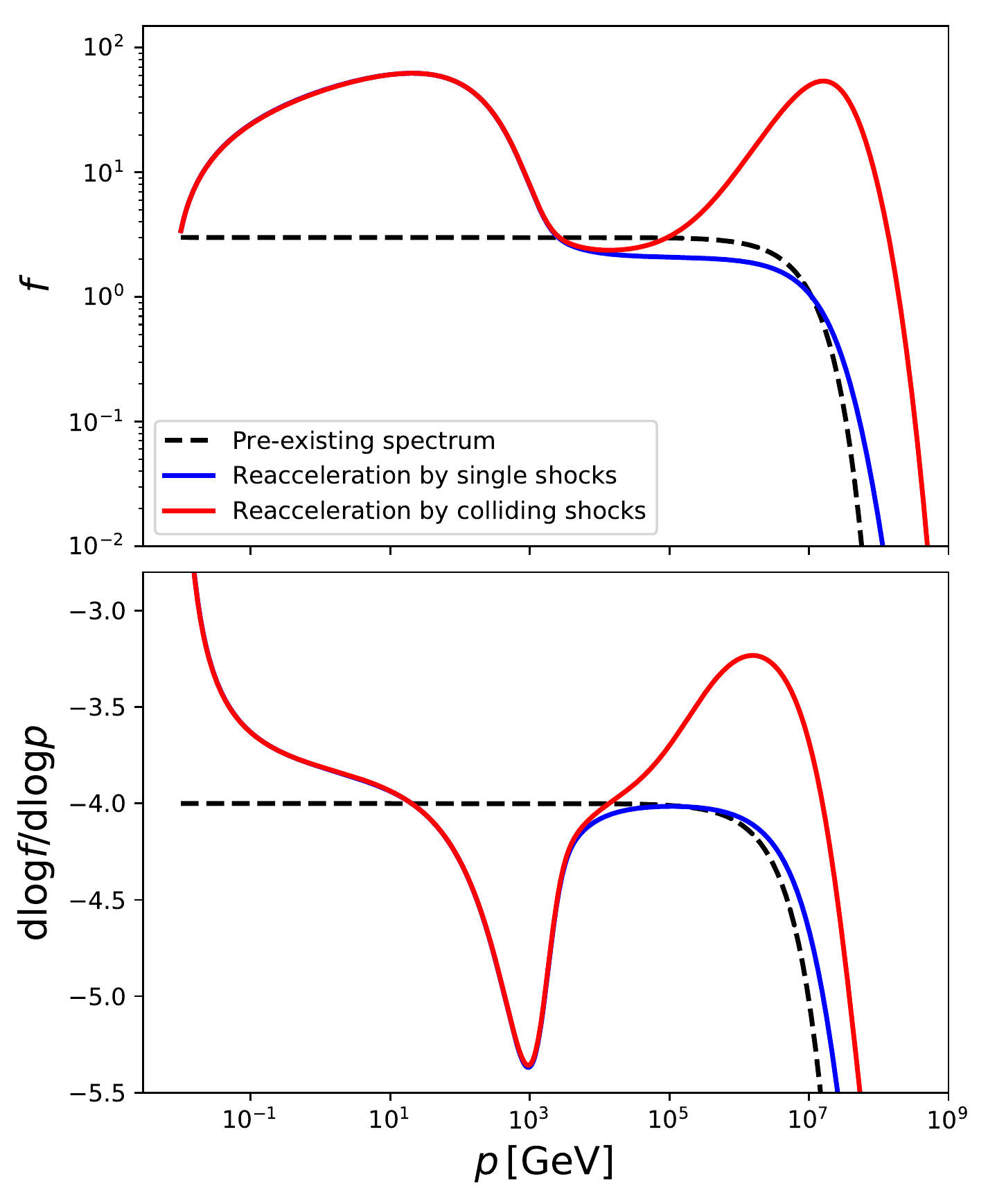}
\caption{\label{fig:reaccelerated_spectrum}Spectrum reaccelerated by colliding shocks in comparison with reacceleration by single shocks.}
\end{figure}

The spectrum obtained from the reacceleration by colliding shocks (also including a fresh acceleration component) is plotted in red in the top panel of Figure \ref{fig:reaccelerated_spectrum}, while the blue line shows the reacceleration by two isolated shocks which last the same time as the collision time of the converging shocks. The bottom panel displays the corresponding spectral indices.

The system of colliding shocks pushes the particles of low energies (0.1 - 1 GeV) towards intermediate energies (10 - 100 GeV), up to the maximum energy which can be reached in the reacceleration process (limited by the finite acceleration time inherent to the time-dependency of the system of converging shocks). This leads to a well-known spectral hardening \citep{melrose1993,cristofari2019}.

The maximum energy of the first bump (around 100~GeV) is due to the finite acceleration time of the system of colliding flows: above its maximum energy, the energy of the preexisting particles is not increased efficiently anymore (their energy gain becomes negligible compared to their initial energy).

The second hardening starting around 10~TeV is due to the collective effect of the two shocks, which
redistribute the high-energy particles in order to tend towards the asymptotic solution $f(p)\sim p^{-3}$ (see Sect.~\ref{sec:hardening}). However, this asymptotic solution is not reached in our example, because the maximum energy of preexisting cosmic rays has been set to 10~PeV.

The resulting particle distribution is similar to the recently modeled spectrum of cosmic-ray leptons escaping from the bow shock pulsar wind nebula of the millisecond pulsar PSR~0437--4715 \citep{bykov2019}. This system is thought to be a site of efficient particle reacceleration in the colliding shock flow zone between the pulsar termination shock and the bow shock.

Although we show that reacceleration effects may be observable at high energies in physical systems where converging shocks appear, we point out that an important hypothesis must be made: that is, preexisting particles need to remain confined between the two converging shocks until their collision. This is not the case if, for instance, the colliding shocks have a much smaller spatial extension than the confining system. As the issue of the confinement introduces a new scale in the problem, another quantitative analysis must be performed. Qualitatively, one expects that beyond the maximum energies that the second accelerator can confine, the preexisting spectrum $f(p)\sim p^{-4}$ should be recovered, while below this limit our results would still be valid. Whether this would erase the hard component at high energies or not depends on the properties of the accelerators and is an open question left for future work.

Finally, we would like to remind that the asymptotic index towards which the reaccelerated spectrum tends at the highest energies is not necessarily $-3$. In a more general context, it is given by Equation \ref{asymptoticindex}. For instance, if the shocks are slowing down and some magnetic field amplification $B\sim V$ takes place, one expects this index to be around -3.4.

\section{An analogy with standing shocks}
\label{sec:standing}

Before concluding, in this Section we highlight a mathematical analogy between the self-similar transport Equation \ref{selfsimilar_transport_Equation} and the transport equation describing a system of \textit{stationary} standing shocks.

Consider a converging flow where matter moves along the x axis with a velocity $\pm u_1$. At the positions $x = \pm x_s$ matter will flow across two stationary shocks. Such a configuration was considered by \citet{siemieniec2000} to describe the accretion of matter onto cosmological sheets, such as the local Supercluster plane. It could also describe in an approximate manner the region of collision of two stellar winds \citep[see e.g.][]{reimer2006}.

In the \textit{downstream} region between the two shocks \citet{siemieniec2000} considered a linear profile for the velocity, $u_2(x) = - u_1 x / (4 x_s)$. 
Following \citet{reimer2006} we also add, between the two shocks, a sink term for particles. This might be due to an advective motion of the shocked plasma along a direction orthogonal to the x axis \citep[see][for further details]{reimer2006,grimaldo2019}. For simplicity, we assume the sink term to be proportional to the particle number density. Under these conditions the stationary transport equation for accelerated particles between the two shocks can be written in a way which is very similar to Equation \ref{selfsimilar_transport_Equation}:
\begin{equation}
- X \d_X f - \d_X [K \d_X f] = - \frac{P}{3} \d_P f + \tilde{Q} \delta(X \pm 1) \delta (P-1) - 4 \frac{f}{\theta}~, \end{equation}
where $X = x/x_s$, $K = 4 \kappa /(x_s u_1)$, $\tilde{Q} = 4 Q /(u_1 p_i)$, $\theta = \tau u_1/x_s$. Here, $\theta$ accounts for particle escape, $\tau$ being an energy independent escape time.


Assuming a homogeneous diffusion coefficient and a power-law spectrum $P \d_P f = - 4 \beta f$, the above equation can be solved to give the particle distribution function in the downstream region:
\begin{equation}
    f_d =~_1F_1 \(\frac{2\beta}{3} - \frac{2}{\theta};\frac{1}{2};-\frac{X^2}{2 K} \) {f_0} \sim f_0 \( 1 + \( \frac{2}{\theta} - \frac{2\beta}{3} \) \frac{X^2}{K} \) ~,
\end{equation}
where $_1F_1$ indicates the confluent hypergeometric function of the first kind.

Upstream, the velocity profile being constant, the transport equation reads:
\begin{equation}
    f_u - K/4 \d_X f_u = 0~, 
\end{equation}
which admits the standard solution:
\begin{equation}
    f_u = f_s \exp{\( \frac{4}{K} (X+1) \)} \, .
\end{equation}
Finally, flux conservation at each shock surface reads:
\begin{equation}
    - [ K \d_X f]_{-1^-}^{-1^+} = 4 \beta f_s + \tilde{Q} \delta (P-1)~, 
\end{equation}
Putting this altogether, one obtains for $K \gg 1$ (high energy limit):
\begin{equation}
    - 4 \beta = - 3 \( 1 + \frac{1}{\theta} \) ~. 
\end{equation}

The solution shows that the spectral index -3 is recovered only in the limiting case where the sink term is zero. This is the hardest possible spectrum on can get, being obtained in the limit of infinite escape time $\tau$ (or $\theta$). The solutions obtained by \citet{siemieniec2000} and \citet{bykov2013} belong to this category.
For finite values of $\theta$ the spectrum is always steeper than $p^{-3}$, and the spectral slope increases monotonically as $\theta$ decreases.


\section{Conclusions}
\label{sec:conclusions}
In this paper we have described the time-dependent problem of particle acceleration in a system of colliding shocks using a self-similarity hypothesis validated numerically. The resulting spectrum of accelerated particles was found to follow quite closely the standard $f \sim p^{-4}$ result, except in the limit of late times and high energies, where a spectral hardening is expected. 
However we showed that this hardening is only asymptotic and not realised when a more realistic time-dependent setup of the problem is considered.
This is due to the finite time available before the shock collision, which does not allow the formation of the asymptotic solution provided by earlier claims.
Therefore, the only impact induced by the combined effects of the two converging shocks is a broadening of the spectral cut-off and the appearance, for some choice of parameters, of a bumpy feature at the highest particle energies. Finally, the maximum energy is not expected to increase by more than a factor of two compared to the case of acceleration at a single shock \cite[see also][]{axford1990}.

The mathematical formalism developed here can also be used to model particle acceleration at a couple of standing shocks in converging flows (e.g. accretion of matter on a cosmic structure, or wind-wind collisions). In this case, we showed that the resulting particle spectra are power laws in energy, with a slope that depends on how effectively particles can escape the acceleration region from downstream.
When no escape is allowed, one recovers the asymptotic solution $f \sim p^{-3}$, which is the hardest possible solution of the problem.


Finally, our semi-analytical framework allowed us to investigate the possibility of particle reacceleration at colliding shocks. We showed that this could result in a pronounced spectral hardening at high energies, but only in situations where the escape of particles from the upstream of the system is strongly suppressed.


\section*{Acknowledgements}
The authors acknowledge support from Agence Nationale de la Recherche (grant ANR- 17-CE31-0014). SG acknowledges support from the Observatory of Paris (Action F\'ed\'eratrice CTA). TV acknowledges Alexandre Marcowith and Etienne Parizot for helpful discussions.




\bibliographystyle{mnras}
\bibliography{biblio} 



\appendix

\section{The general form of the self-similar distribution function}
\label{app1}

In this Appendix we show that Equation \ref{selfsimilar_transport_Equation} is very general and can account for a large variety of phenomena.

\subsection{Time-dependent shock velocity}

First, we do not assume a constant velocity for the movements of the shocks, but rather a time-dependent one $V(t)$. The situation is still symmetric but one could for instance account for some deceleration due to the backreaction of the fluid. In this case, our previous rescaling arguments are still expected to be valid, except that $L-Vt$ should be replaced by $L-\int_0^t \dd t' V(t') = L-t \langle V \rangle _t \equiv \frac{1}{t}$. Here and in the following, to simply notations we define:
\begin{equation}
    \langle V \rangle _t \equiv \frac{1}{t} \int_0^t \dd t' V(t')~.
\end{equation}

\subsection{Generalisation of the diffusion coefficient}

Our second generalisation concerns the turbulence spectrum. We now assume that the diffusion coefficient takes the following form:
\begin{equation}
    \kappa = \kappa_0 (t) p^\delta~.
\end{equation}
We have introduced a time dependency through $\kappa_0(t)$ which can for instance account for a phenomenological magnetic field amplification, and a generic turbulence index $\delta$.

The self-similar hypothesis is then generalised as:
\begin{align}
\label{app2step1}
    f(x,p,t) = \lambda^{\beta} \( \frac{\kappa_0 (t)}{\kappa_0 (t_0)}\)^{\gamma} f \( \lambda x,\( \lambda \frac{\kappa_0 (t)}{\kappa_0 (t_0)}\)^\frac{1}{\delta} p , t_0 \)~,
\end{align}
where $\lambda \equiv (L-\langle V \rangle_{t_0} t_0)/(L-\langle V \rangle_{t} t)$.

Taking the time derivative of Eq.~\ref{app1step1} (see App.~\ref{app2}) we obtain:
\begin{multline}
    \d_t f(x,p,t) - \frac{\d_t \lambda}{\lambda} x \d_x f(x,p,t)  = \\
       \(\beta \frac{\d_t \lambda}{\lambda}  + \gamma \frac{\d_t \kappa_0(t)}{\kappa_0(t)} \) 
    f(x,p,t) +
    \( \frac{\d_t \lambda}{\lambda} + \frac{\d_t \kappa_0(t)}{\kappa_0(t)} \) \frac{p}{\delta} \d_p f(x,p,t)~.
\end{multline}
We assume that in the limit of small times and small energies the downstream spectrum is a stationary homogeneous power-law of spectral index $s$. This implies:
\begin{equation}
    \(\frac{s}{\delta} - \gamma\) \frac{\d_t \kappa_0(t)}{\kappa_0(t)}  = \(\beta - \frac{s}{\delta}\) \frac{\d_t \lambda}{\lambda}~.
\end{equation}
And therefore, by integrating we find that in order to be ``self-similar friendly'', the time-dependency of the diffusion coefficient must read:
\begin{equation}
    \kappa_0(t) = \kappa_0(t_0) \lambda^{\zeta-1} \, , \qquad \zeta \equiv \frac{(\beta-\gamma)\delta}{s-\gamma \delta}~.
\end{equation}
Then, when replaced in \ref{app2step1}, this gives:
\begin{align}
\label{generalised_self_similar_hyp}
    f(x,p,t) = \lambda^{s \frac{\zeta}{\delta}} f \( \lambda x,\lambda^\frac{\zeta}{\delta} p , t_0 \)~,
\end{align}
such that for $s=4$ (strong shocks), one recovers Equation~\ref{selfsimilar_transport_Equation} (setting $\alpha \equiv s \zeta/\delta$)

\section{Modification of the spectral index in converging shocks}
\label{app2}
We start from Eq.~\ref{selfsimilar_transport_Equation} with a generic power-law spectrum ($P \d_P f = -s \beta f$) and we set $\xi \equiv (1 - \beta) \alpha$ to simplify notation:
\begin{equation}\label{app1step1}
\begin{split}
\xi f + (X + U) \d_X f - \tilde{K} \d^2_{X} f = - \frac{s \beta}{3} \d_{X} U f
\end{split}
\end{equation}
The injection has been omitted as the aim here is to probe high momenta.

The solution upstream ($U=0$) is:
\begin{equation}
\begin{split}
    f_u = \, _1F_1 \(\frac{\xi}{2};\frac{1}{2};\frac{X^2}{2 \tilde{K}} \) {f_0} \underset{\tilde{K} \gg 1}{\sim} {f_0} \(1 + \xi \frac{X^2}{2 \tilde{K}} \)~,
\end{split}
\end{equation}
while the solution downstream ($U \neq 0$) is:
\begin{equation}
\begin{split}
    f_d = {\mathcal{C}_1} U \(\frac{\xi}{2},\frac{1}{2},\frac{(U+X)^2}{2 \tilde{K}}\)
    + \mathcal{C}_2 \, _1F_1\(\frac{\xi}{2};\frac{1}{2};\frac{(U+X)^2}{2 \tilde{K}}\)
    \\ \underset{\tilde{K} \gg 1}{\sim}
    \mathcal{C}_1' \(1 + \frac{\xi (U+X)^2}{2 \tilde{K}} \) + \mathcal{C}_2' \frac{(U+X) }{\sqrt{\tilde{K}}} \frac{\Gamma \(\frac{1+\xi}{2}\)}{\Gamma \(\frac{\xi}{2} \)~, }
\end{split}
\end{equation}
where $U(a,b,x)$ is the Tricomi confluent hypergeometric function.

This allows to compute the asymptotic difference of the derivatives around the shock at $X=-1$:
\begin{equation}
    -\left[ \d_X f_u - \d_X f_d \right]_{X=-1} \sim
    \frac{\xi f_0}{\tilde{K}} U + \frac{ \mathcal{C}_2' \Gamma \(\frac{\xi +1}{2}\)}{\sqrt{\tilde{K}} \Gamma \(\xi/2\)}~,
\end{equation}
where we have imposed the continuity of $f$ around the shock to express the constant $\mathcal{C}_1'$ as function of $\mathcal{C}_2'$ and $f_0$.

Then, by integrating Equation \ref{app1step1} around the shock we get at first order in $1/\tilde{K}$:
\begin{align}\label{app1step2}
    \frac{\xi f_0}{\tilde{K}} U + \frac{\mathcal{C}_2' \Gamma \(\frac{\xi +1}{2}\)}{\sqrt{\tilde{K}} \Gamma \(\xi/2\)}
    \sim \frac{U}{3 \tilde{K}} s \beta f_{-1}
    \sim \frac{U}{3 \tilde{K}} s \beta {f_0}~,
\end{align}
which is valid for any (big) $\tilde{K}$, hence each order should vanish independently and we obtain:
\begin{align}
\label{asymptoticindex}
    \beta \sim \frac{1}{1+s/(3\alpha)}~.
\end{align}
Assuming the natural scaling exponent $\alpha = 4$ (such that the critical momentum decreases linearly with time), we get $\beta = 3/4$, hence a spectral index of $-3$. Interestingly, the spectrum steepens if $\alpha$ increases. For instance, if above some energy there are non linear phenomena which decrease the typical energy quicker than $L-Vt$, say for instance $\alpha = 8$, then the spectral index is -3.4 and one could expect a broken spectrum (but still hard).


\section{Maximum energy}
\label{app3}
When a relativistic particle crosses a shock, its momentum is increased by $\Delta p = p V/(2 c)$. The typical time between two crossings is the downstream diffusion time plus the minimum between the upstream diffusion time and the time it takes to travel from one shock to the other.
However if the latter is much smaller than the former, then it is negligible compared to the downstream diffusion time. We have therefore three regimes: i/ large distance between the shocks, which implies that the particles wait two diffusion times between crossing the shocks; ii/ the distance between the shocks is of the order of two diffusion lengths; iii/ the distance between the shocks is negligible.
We therefore obtain:
\begin{align}
    \Delta t \propto \frac{\xi \kappa}{V^2} \Delta p / p
\end{align}
where the proportionality constant is 1 for case (i) and (ii) and 1/2 for case (iii), and where $\xi$ is a model-dependent number.
Assuming Bohm diffusion and constant shock velocities we can separate variables and integrate to get:
\begin{align}
    P_m &= \mathcal{I}(t) \, \qquad {\rm for~cases~} (i), \, (ii) \\
    P_m &= P_c + \int_{t_c}^t \frac{2 V}{L K_0 \xi } \dd t'
    = 2 \mathcal{I}(t) - P_c \, \qquad {\rm for~case~} (iii)~,
\end{align}
where $t_c$ is the time at which $P_m = P_c =  (\lambda K_0)^{-1}$, i.e. $t=t_c$ is the time of the transition from regime (ii) to regime (iii). $\mathcal{I}(t)$ is defined by:
\begin{equation}
    \mathcal{I}(t) \equiv 1 + \frac{1}{L K_0 \xi } (t-t_0) V \sim \frac{1}{L K_0 \xi } (t-t_0) V~,
\end{equation}
where we assumed $K_0 \ll 1$. This implies that $t_c$ is solution of:
\begin{equation}
    \frac{L-Vt_c}{(t_c-t_0) V} = \frac{L-V t_0}{L \xi}  \sim 0~,
\end{equation}
where we assumed $\xi \gg 1$. Hence $t_c \sim L/V$ and we actually almost never see regime (iii), such that one can set $P_m(t) \sim \mathcal{I}(t)$ at any time.

\section{Generalised spectrum}
\label{app4}
We assume that the diffusion coefficient scales as:
\begin{equation}
    \kappa(p,t) = \kappa_0(t) p^\delta = \kappa_0(t_0) \lambda^{\zeta-1} p^\delta
\end{equation}
where $\lambda \equiv (L-\langle V \rangle_{t_0} t_0)/(L-\langle V \rangle_{t} t)$, $\langle V \rangle _t \equiv \frac{1}{t} \int_{t_0}^t \dd t' V(t')$.
As shown in Appendix \ref{app1}, the self-similar hypothesis is then formulated as:
\begin{equation}
    f(x,p,t) = \lambda^{s \frac{\zeta}{\delta}} f \( \lambda x,\lambda^\frac{\zeta}{\delta} p , t_0 \)
\end{equation}
And as shown in Appendix \ref{app2}, the spectral index at $\tilde{K}\gg 1$ is:
\begin{equation}
     -s \beta \sim -\frac{s}{1+\delta/(3 \zeta)}
\end{equation}
Setting $V(t) = V_0 \(\frac{t}{t_0}\)^{-\nu}$, The integrated spectrum can be shown to read:
\begin{multline}
\label{greenfunctionfinal}
     S(p_0,p) = \frac{\mathcal{A}'}{p_0} \(\frac{p}{p_0}\)^{-s} \int_0^1 \dd w \( 1 + \(\frac{K_0 p^\delta}{(1-w)^{\zeta}}\)^{\frac{s}{3 \zeta+\delta}} \) \\ \times \exp \( \frac{- \xi K_0 (1-2 \nu) p}{\xi K_0 (1-2 \nu)p_0 + \chi^\frac{\nu}{1-\nu} ((1-\nu)w)^{\frac{1-2\nu}{1-\nu}}-\chi} \) \(\frac{w \chi}{\chi+(1-\nu)w}\)^{\frac{\nu}{1-\nu}}
\end{multline}
which is also, up to a normalisation constant, the Green function of a system of self-similar converging shocks with time-dependent velocity (through the parameters $\chi = V_0 t_0/L$ and $\nu$), amplification phenomena (through the exponent $\zeta$), a generic turbulence spectrum (through $\delta$) and a phenomenological escape.

It is clear from Equation~\ref{greenfunctionfinal} that the exponent $\nu$ is the key parameter to see the peculiar features of the system of converging shocks. Indeed, the higher $\nu$, the bigger importance is given to values of $w$ close to one, thanks to the last factor, and $w$ close to one means late times. Time-integrated spectra for various values of $\nu$ are shown in Fig.~\ref{fig:green_function}. 

\bsp	
\label{lastpage}
\end{document}